\documentclass[]{spie}  %>>> use for US letter paper
\usepackage{xcolor}
\usepackage{amsmath,amsfonts,amssymb}
\usepackage{graphicx}
\usepackage[colorlinks=true, allcolors=blue]{hyperref}
\usepackage{siunitx}
\usepackage{makecell}
\usepackage[acronym]{glossaries}
\usepackage[nolist]{acronym}
\newacro{ADC}[ADC]{Atmospheric Dispersion Corrector}
\newacro{SAF}[SAF]{Science Archive Facility}
\newacro{SOXS}[SOXS]{Son-Of-X-shooter}
\newacro{NTT}[NTT]{New Technology Telescope}
\newacro{ESO}[ESO]{European Southern Observatory}
\newacro{DRAWER}[DRAWER]{Data Reduction And WEllness Reporter}
\newacro{AC}[AC]{Acquisition and Imaging Camera}
\newacro{CPL}[CPL]{Common Pipeline Library}
\newacro{CLI}[CLI]{Command-Line Interface}
\newacro{QC}[QC]{Quality Control}
\newacro{E2E}[E2E]{End-to-End}

\usepackage{color}
\usepackage{colortbl}
\usepackage{multirow}
\usepackage{listings}
\usepackage{float}
\usepackage{color}
\usepackage{xcolor,colortbl}
\usepackage{caption}
\usepackage{subcaption}
\definecolor{red}{HTML}{dc322f}
\definecolor{green}{HTML}{859900}
\definecolor{blue}{HTML}{268bd2}
\definecolor{orange}{HTML}{cb4b16}
\definecolor{cyan}{HTML}{2aa198}
\definecolor{magneta}{HTML}{d33682}
\definecolor{violet}{HTML}{6c71c4}

\definecolor{dkgreen}{rgb}{0,0.6,0}
\definecolor{gray}{rgb}{0.5,0.5,0.5}
\definecolor{mauve}{rgb}{0.58,0,0.82}

\title{The Quality Check system architecture for Son-Of-X-Shooter SOXS}

\author[a,b,c]{Marco Landoni}
\author[e]{Laurent Marty}
\author[d]{Dave Young}
\author[a-b]{Laura Asquini}

\author[d]{Stephen J. Smartt}

\author[a]{Sergio~Campana}
\author[f]{Riccardo~Claudi}
\author[e]{Pietro~Schipani}
\author[a]{Matteo~Aliverti}
\author[f]{Federico Battaini}
\author[f]{Andrea~Baruffolo}
\author[g]{Sagi~Ben-Ami}
\author[h]{Federico~Biondi}
\author[a]{Andrea Bianco}
\author[e]{Giulio~Capasso}
\author[k]{Rosario~Cosentino}
\author[i]{Francesco~D'Alessio}
\author[a]{Paolo~D'Avanzo}
\author[a]{Matteo~Genoni}
\author[g]{Ofir	Hershko}
\author[m]{Hanindyo~Kuncarayakti}
\author[k]{Matteo~Munari}
\author[n]{Giuliano~Pignata}
\author[s]{Adam~Rubin}
\author[k]{Salvatore~Scuderi}
\author[i]{Fabrizio~Vitali}
\author[l]{Jani~Achrén}
\author[n]{José~Antonio~Araiza-Duran}
\author[o]{Iair~Arcavi}
\author[p]{Anna~Brucalassi}
\author[g]{Rachel~Bruch}
\author[f]{Enrico~Cappellaro}
\author[e]{Mirko~Colapietro}
\author[e]{Massimo~Della~Valle}
\author[e]{Marco~De~Pascale}
\author[k]{Rosario~Di~Benedetto}
\author[e]{Sergio~D'Orsi}
\author[g]{Avishay~Gal-Yam}
\author[q]{Marcos~Hernandez}
\author[m]{Jari~Kotilainen}
\author[i]{Gianluca~Li~Causi}
\author[m]{Seppo~Mattila}
\author[a]{Luca Oggioni}
\author[a]{Giorgio Pariani}
\author[g]{Michael~Rappaport}
\author[f]{Kalyan~Radhakrishnan}
\author[f]{Davide~Ricci}
\author[a]{Marco~Riva}
\author[f]{Bernardo~Salasnich}
\author[k]{Ricardo~Zanmar~Sanchez}
\author[r]{Maximilian~Stritzinger}
\author[q]{Hector~Ventura}

\affil[a]{INAF – Osservatorio Astronomico di Brera-Merate, via E. Bianchi 46, I-23807 Merate (LC), Italy;}
\affil[b]{Dipartimento di Scienza e Alta Tecnologia, Università dell’Insubria, via Valleggio 11, I-22100 Como, Italy}
\affil[c]{INAF - Osservatorio Astronomico di Cagliari. Via della Scienza 5, Selargius (CA) - Italy}
\affil[d]{Astrophysics Research Centre, School of Mathematics and Physics, Queen's University Belfast, Belfast BT7 1NN, UK}
\affil[e]{INAF - Osservatorio Astronomico di Capodimonte, Salita Moiariello 16, Naples- Italy }
\affil[f]{INAF -- Osservatorio Astronomico di Padova, Vicolo dell’Osservatorio 5, I-35122, Padua, Italy }

\affil[g]{Weizmann Institute of Science, Herzl St 234, Rehovot, 7610001, Israel }
\affil[h]{Max-Planck-Institut für Extraterrestrische Physik, Giessenbachstr. 1, D-85748 Garching, Germany }
\affil[k]{INAF -- Osservatorio Astrofisico di Catania, Via S. Sofia 78 30, I-95123 Catania, Italy }
\affil[i]{INAF -- Osservatorio Astronomico di Roma, Via Frascati 33, I-00078 M. Porzio Catone, Italy }
\affil[l]{Incident Angle Oy, Capsiankatu 4 A 29, FI-20320 Turku, Finland }
\affil[m]{Finnish Centre for Astronomy with ESO (FINCA), FI-20014 University of Turku, Finland}
\affil[n]{Universidad Andres Bello, Avda. Republica 252, Santiago, Chile }
\affil[o]{Tel Aviv University, Department of Astrophysics, 69978 Tel Aviv, Israel }
\affil[p]{INAF - Osservatorio Astrofisico di Arcetri.Largo Enrico Fermi 5, 50125 Florence - ITALY }

\affil[q]{FGG-INAF, TNG, Rambla J.A. Fernández Pérez 7, E-38712 Breña Baja (TF), Spain }
\affil[r]{Aarhus University, Ny Munkegade 120, D-8000 Aarhus, Denmark }
\affil[s]{ESO – European Southern Observatory
Karl-Schwarzschild-Straße 2, 85748 Garching bei München, Germany}
\authorinfo{Contact authors: Laurent Marty (laurent.marty@inaf.it) and Marco Landoni (marco.landoni@inaf.it)}
 
\begin{document} 
\maketitle

\begin{abstract}
We report the implemented architecture for monitoring the health and the quality of the Son Of X-Shooter (SOXS) spectrograph for the New Technology Telescope in La Silla at the European Southern Observatory. Briefly, we report on the innovative no-SQL database approach used for storing time-series data that best suits for automatically triggering alarm, and report high-quality graphs on the dashboard to be used by the operation support team. The system is designed to constantly and actively monitor the Key Performance Indicators (KPI) metrics, as much automatically as possible, reducing the overhead on the support and operation teams. Moreover, we will also detail about the interface designed to inject quality checks metrics from the automated SOXS Pipeline (Young et al. 2022).
\end{abstract}

\keywords{SOXS, noSQL, Quality control, Quality Checks, Pipeline, Data Reduction, Spectroscopy, Imaging}

\section{INTRODUCTION}
\label{sec:intro}  % \label{} allows reference to this section

The \ac{SOXS} (Son Of X-Shooter) instrument is a new medium resolution spectrograph ($R\simeq4500$) able to simultaneously observe 350-2000nm (U- to H-band) to a limiting magnitude of R $\sim 20$ (3600sec, S/N $\sim$ 10). It shall be hosted at the Nasmyth focus of the \ac{NTT} at La Silla Observatory, Chile (see \cite{Schipani20} for an overview). This paper describes the design of the \ac{SOXS} quality check and data-flow system. Details of each of the other \ac{SOXS} subsystems can be found in a set of related papers , see \cite{Aliverti18,Aliverti20,Biondi18,Biondi20,Brucalassi18,Brucalassi20,Capasso18,Claudi18,Claudi20,Colapietro20,Cosentino18,Cosentino20,Genoni20,Kuncarayakti20,Ricci18,Ricci20,Rubin18,Rubin20,Sanchez18,Sanchez20,Schipani16,Schipani18,Schipani20,Vitali18,Vitali20,Young20,Marty22,Landoni22}.
A peculiarity of this instrument, as described in Asquini et al 2022 SPIE in press,  the instrument will be operated in the ESO La Silla-Paranal Observatory operation environment without an astronomer on the mountain. From this stems the need to design and develop a suite of software that can autonomously organize and manage the night both in advance (for scientists to approve and scrutinize) and to monitor the health of the instrument. In this paper we focus on the Quality Check System architecture that has been implemented for SOXS.  

\section{The \ac{SOXS} Pipeline - The first block for the Quality control}

\label{sec:remit} 

\begin{figure} [h!]
    \centering
    \includegraphics[width=0.45\textwidth]{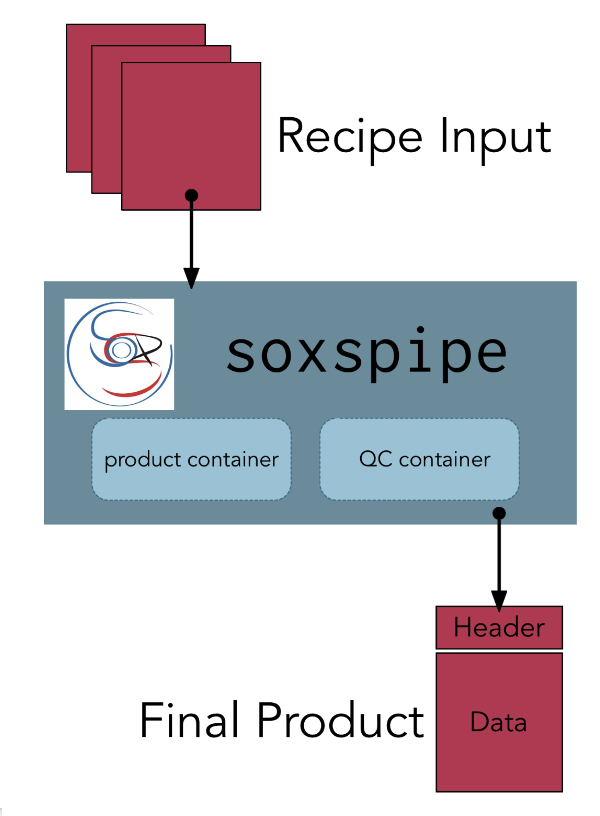}
    \caption{General flow from the soxspipe and the QC container. The QCs computed by the pipeline are stored in the proper fits header of each calibration, intermediate and final product. A daemon, running on the pipeline machine, will read the header and push documents into the no-SQL database collections.}
    \label{fig:overallflow}
\end{figure}

As described in Young et al 2022 SPIE in press, the main purpose of the \ac{SOXS} Data Reduction pipeline is to use \ac{SOXS} calibration data to remove all instrument signatures and deliver Phase 3 compliant science data products. The pipeline also measures parameters on calibration product and intermediate product to assess both the quality of the data and the health of the instrument. The schema shown in Figure \ref{fig:overallflow} reports how the SOXS pipeline interacts with Quality Check Systems by using FITS headers.

\section{A no-SQL approach database}
\label{sec:architecture} 
\begin{figure} [h!]
    \centering
    \includegraphics[width=0.95\textwidth]{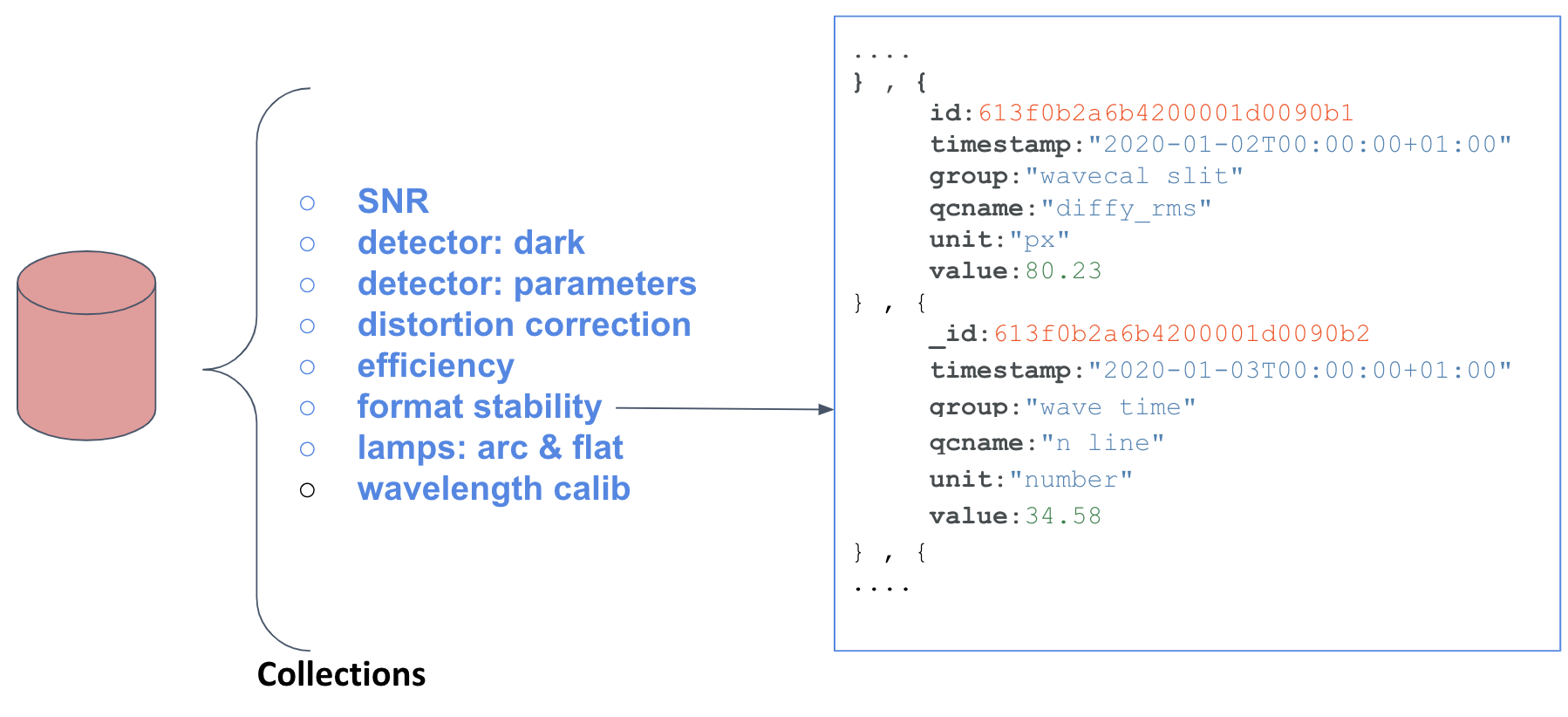}
    \caption{MongoDB collection and document organisation. See Section 2 in the text for details}
    \label{fig:json}
\end{figure}

\begin{figure} [h!]
    \centering
    \includegraphics[width=0.95\textwidth]{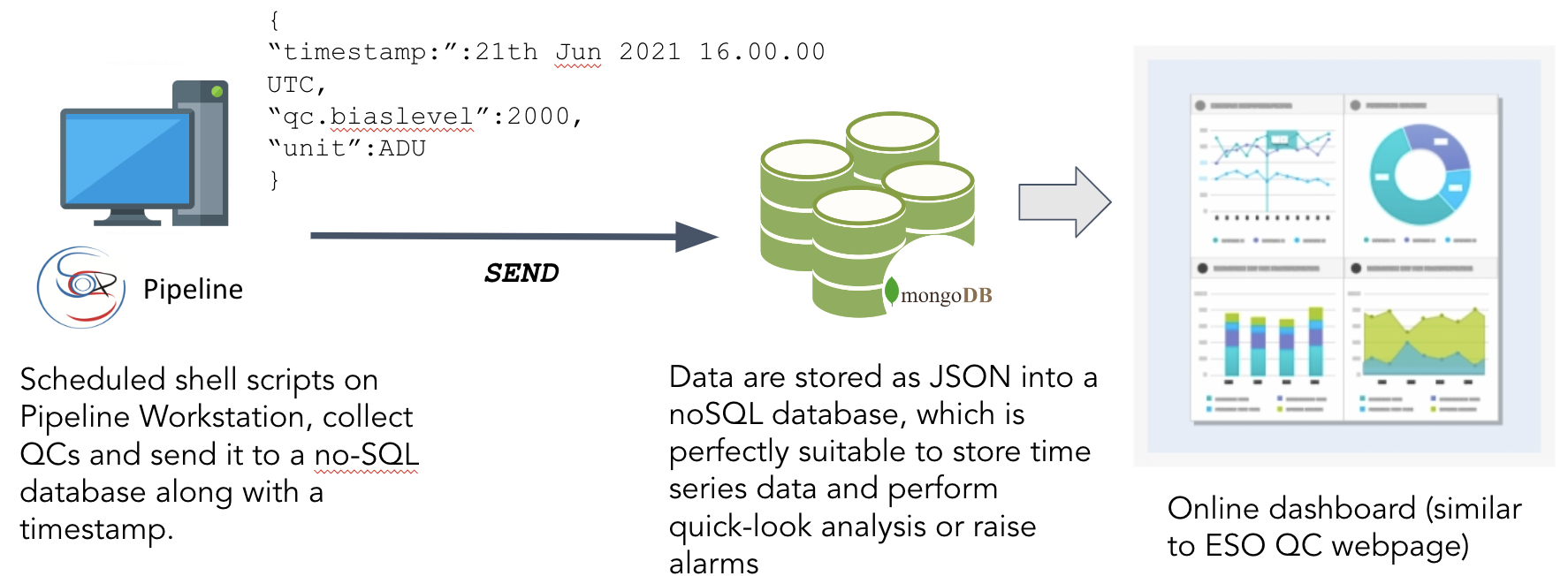}
    \caption{Injection of the data in the MongoDB database. A deamon, running on the pipeline machine, checks for new calibration, intermediate and final product on the filesystem. If new files are found, then the header is parsed and a JSON document is sent to the MongoDB database cluster in the proper collection for storage.}
    \label{fig:injection}
\end{figure}

The Quality Check control for SOXS will store time-series data that will be presented to the user in order to produce both (a) a first glance table that summarises the status of instrument, by comparing measured values on the images with those expected by design (RON, dark,..), and (b) provide interactive plots that can be used to better understand trends, outliers, etc.
\\
\begin{figure} [h!]
    \centering
    \includegraphics[width=0.45\textwidth]{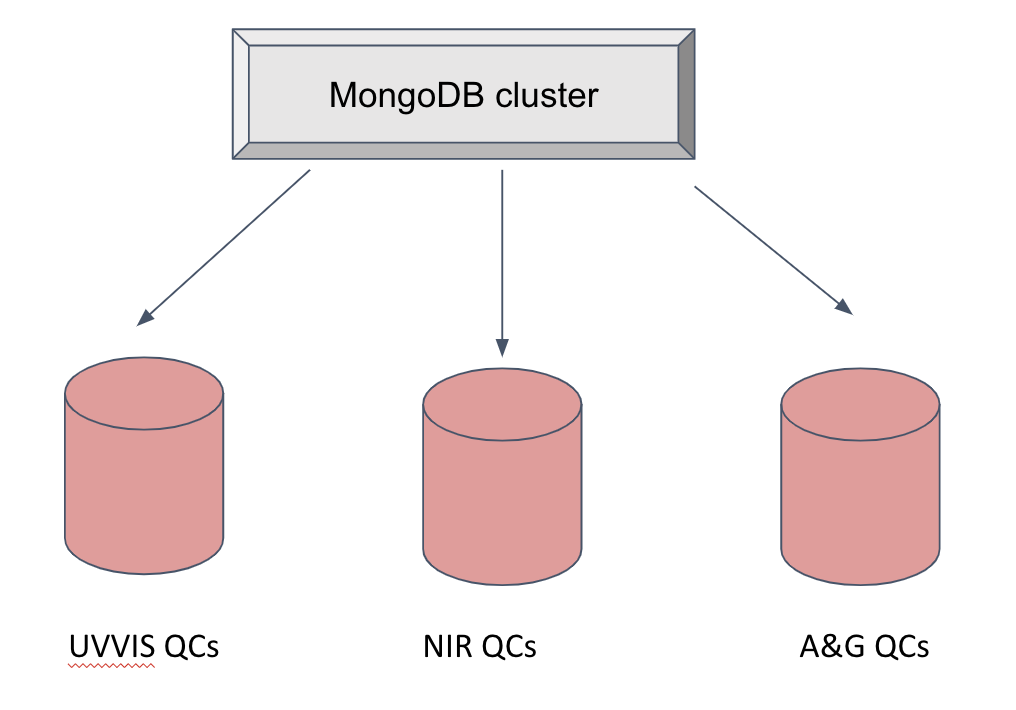}
    \caption{The SOXS quality check system database architecture.}
    \label{fig:cluster}
\end{figure}
Since the structure of the database, the number of monitored Quality Checks and the data structure in general will be fluid both during the design of the instrument and during the operations we decided to adopt a MongoDB no SQL database to store our data points.
The noSQL databases have the great advantage to be fast against certain queries and flexible with respect to the underlying data structure (no a-priori design of the data structure is needed). They basically store each datum as a JSON file in sets called collections. We exploit this to fact to design the overall architecture of our databases as reported in Figure \ref{fig:cluster}.
\\
Briefly, for each arm of the instrument we define in our MongoDB cluster a database that contain JSON documents that are organized in Collections as in the sample of Figure \ref{fig:json}.
For each group of Quality Checks that we intend to maintain under control (dark current, efficiencies, Signal-to-Noise ratio reached, etc.) we define a MongoDB Collection (which is no more than a set of document) to store a JSON document for each measured data-point which contain the measured value with proper unit and a set of metadata used to identify the datapoint. Those metadata are the qcname (a mnemonic label that will be displayed on the Web Application) and the group name which the quality check belongs to. For example, as shown in Figure \ref{fig:json} in the Collection \textit{format stability} we will measure a number wavelength calibration quality checks in the group called \textit{wavecal\_slit}. The quality check value \textit{diffy\_rms} which measures the rms on the wavelenght calibration belongs to the collection \textit{Format Stability} in the group \textit{wavecal\_slit}.Finally, for keeping the data up-to-date a deamon, running on the pipeline machine, checks
for new calibration, intermediate and final product on the filesystem. If new files are found, then the header is
parsed and a JSON document is sent to the MongoDB database cluster in the proper collection for storage (see Figure \ref{fig:injection}).    

\section{THE WEB APPLICATION}
In order to present data to the end-user and to the operation team, a proper Web Application with state-of-the art responsive tecnologies has been developed.
\\

\begin{figure} [h!]
    \centering
    \includegraphics[width=1.1\textwidth]{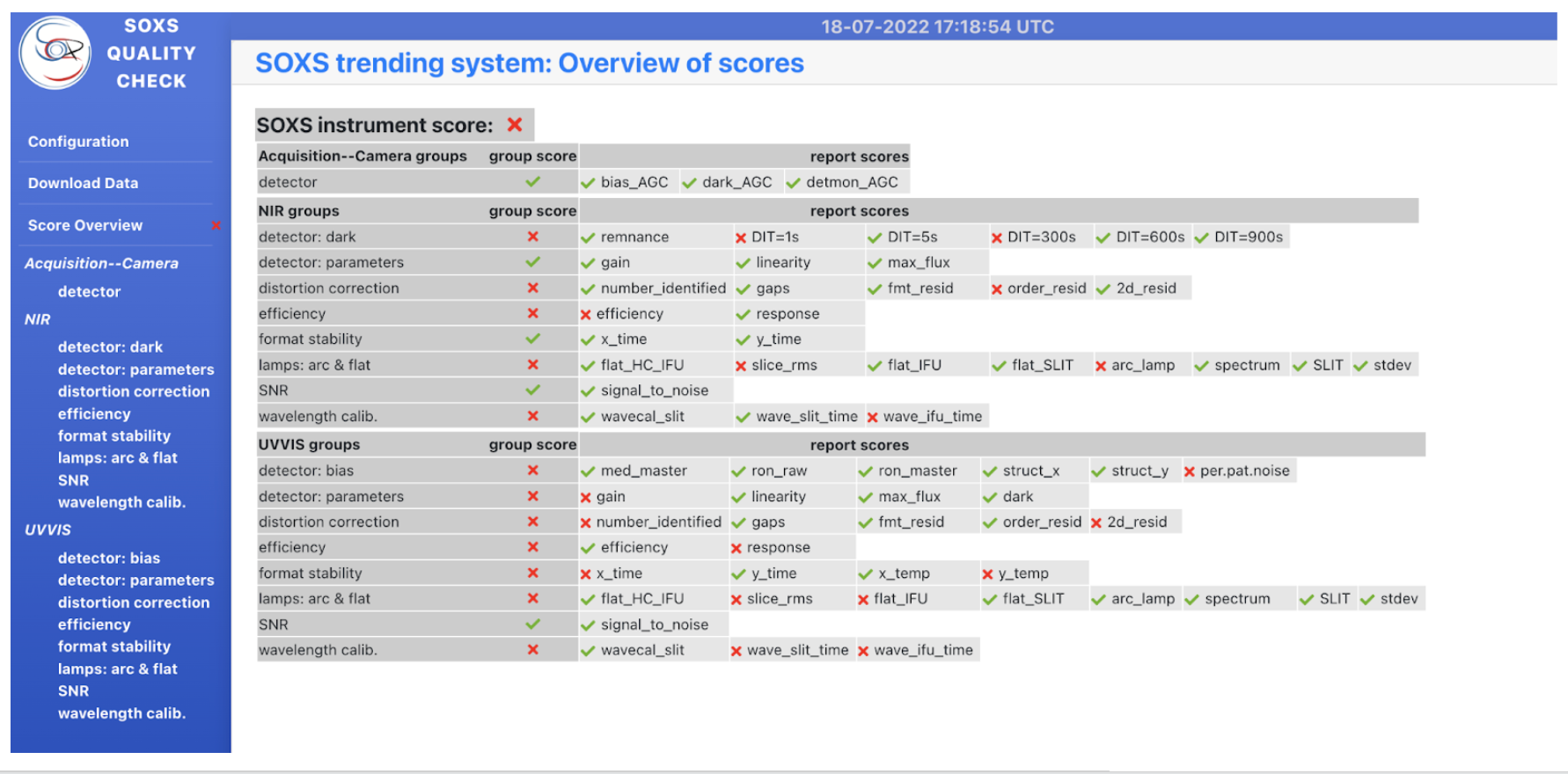}
    \caption{Dashboard of the Web Application for SOXS Quality Checks monitor. See text for details.}
    \label{fig:dashboard}
\end{figure}

\begin{figure} [h!]
    \centering
    \includegraphics[width=1.1\textwidth]{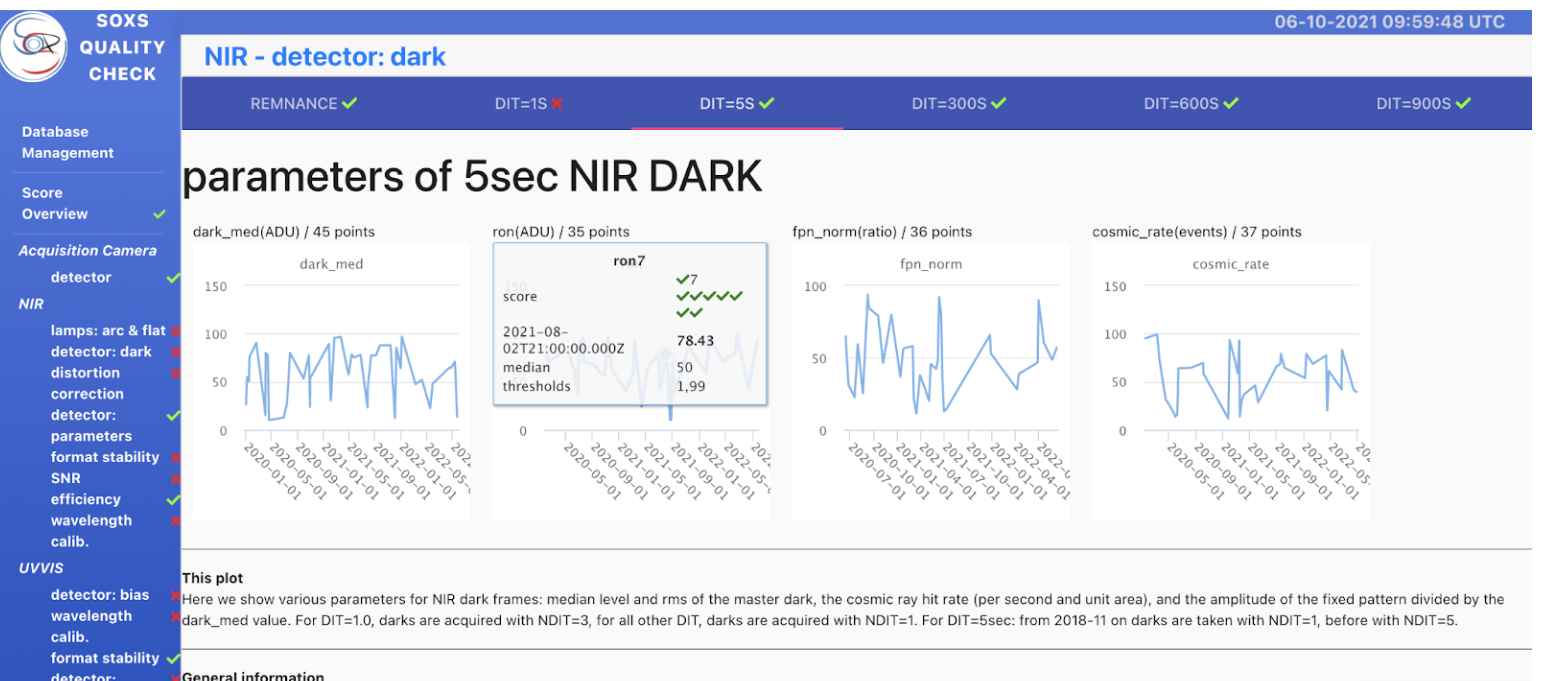}
    \caption{Graphs for the dark current quality checks monitored for SOXS.}
    \label{fig:graph1}
\end{figure}

\begin{figure} [h!]
    \centering
    \includegraphics[width=1.1\textwidth]{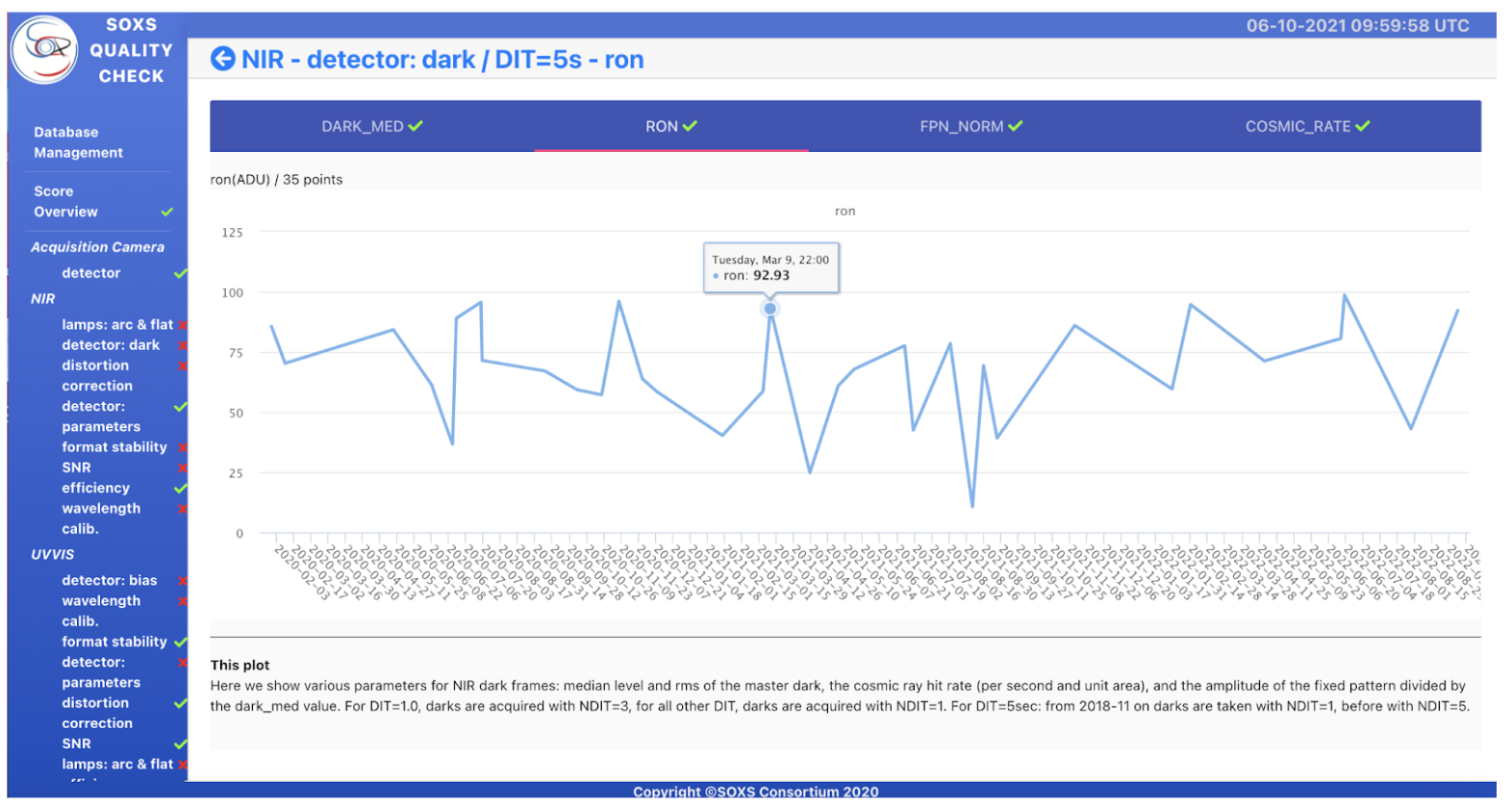}
    \caption{A zoomed view of a graph for a quality check. See text for details}
    \label{fig:graph2}
\end{figure}

The application is in charge to present a Dashboard (see Figure \ref{fig:dashboard}) where the global health status of the instrument is presented. If the series of monitored quality check parameters is OK, a green tick is reported while a red mark is shown when parameters are 1-sigma out of the expected nominal value coming from design (or actual measurements during the Commissioning Phase in La Silla).
\\
The Web application is also capable to show, for each monitored quality check, a detailed and interactive graph that could be used by the operation team to debug problems on the instrument or to monitor trends on relevant parameters of the detector (such as the dark current, the fixed pattern noise, etc.). Figure \ref{fig:graph1} and Figure \ref{fig:graph2} report a detailed view of the graph available for each quality check. The user, by moving the cursor on the plot line, could read the values and their goodness indicated by green or red marks.

The application allows also to download the data, selecting date range and relevant quality checks, in CSV or TXT format for offline analyses with other third-party tools if needed.

\section{ALARMING}
An important part of the monitor system of SOXS is the automatic alarm feature present on the system. In particular, by adopting Cloud Based services offered by Amazon Web Services (see e.g. \cite{landoniaws} ) the system is able to notify the operation team of a faulty monitored quality check by email or SMS. This methodology, coupled with the presented web-based dashboard, allows a very quick instrument monitoring by the operation team both in Europe and Chile and to discover as soon as possible drifting key parameters of SOXS in advance.  
%\section{REFERENCES}
\newpage
% References
\bibliography{report} % bibliography data in report.bib
\bibliographystyle{spiebib} % makes bibtex use spiebib.bst

\end{document}